Mid-infrared dual comb spectroscopy via continuous-wave optical parametric oscillation

D. A. Long[1,*], G. C. Mathews[2], S. Pegahan[3], A. Ross[3], S. C. Coburn[2], P.-W. Tsai[3],

G. B. Rieker[2], A. T. Heiniger[3,*]

[1]National Institute of Standards and Technology, Gaithersburg, MD, USA

[2]Precision Laser Diagnostics Laboratory, University of Colorado, Boulder, CO, USA

[3]TOPTICA Photonics, Pittsford, NY, USA

[*]Corresponding authors: david.long@nist.gov (D. A. Long) and adam.heiniger@toptica-usa.com

**Abstract**

Dual-comb spectroscopy has demonstrated remarkable capabilities for rapid and sensitive measurements; however, significant challenges still exist in generating high-power, mutually coherent mid-infrared combs. Here we demonstrate that a pair of near-infrared femtosecond frequency combs can be spectrally translated via a *continuous-wave* optical parametric oscillator. The pair of spectrally translated combs demonstrated high mutual coherence, power per comb tooth in excess of hundreds of microwatts, and were tunable between 4 µm and 5 µm. Unlike previous approaches which relied upon synchronous optical parametric oscillation, the present approach avoids challenges associated with comb stabilization, low power per comb tooth, and complex cavity designs. Further it is readily amenable to high repetition rates (gigahertz-level and beyond). The flexible and facile nature of this approach provides a robust path for the spectral translation of mode-locked combs, achieving spectral bandwidths limited only by the phase matching bandwidth of the optical parametric oscillator. This approach holds significant promise for applications in chemical kinetics, remote sensing, combustion science, and precision spectroscopy, where the combination of high powers, broad bandwidths, and high measurement rates are transformative.

**Introduction**

Dual comb spectroscopy with mode-locked lasers has emerged as a transformational tool for chemical and physical science research due to its unprecedented combination of wide spectral bandwidths, high resolution, rapid acquisition rate, and high sensitivity[1,2]. While initial demonstrations were predominantly conducted in the readily accessible near-infrared spectral region, recent efforts have focused on extending these approaches to the mid-infrared[3-12] and ultraviolet regions[13-15]. Spectral translation to the so-called fingerprint region in the mid-infrared is critical to reach strong, distinct molecular absorption features, allowing for the detection of complex species with high sensitivities and measurement rates. Generally, spectral translation of a mode-locked comb to the mid-infrared is performed via either difference frequency generation (DFG)[4,6,9] or optical parametric oscillation (OPO)[3,8,16-20]. OPO is a particularly promising method as it offers the potential for high efficiency, wide tuning ranges, and very high optical powers.

Recently, we demonstrated that a pair of near-infrared electro-optic frequency combs can be spectrally translated by a continuous-wave OPO since those combs were fundamentally continuous wave in nature[21]. Although this approach was capable of achieving nanosecond time resolution dual-comb spectroscopy, the combs themselves were limited to just 14 comb teeth. Previous work has shown that both counter-propagating[18,19] and co-propagating,[16,17] singly resonant OPOs can be synchronously pumped with mode-locked lasers to produce mid-infrared dual combs. Despite these efforts, pumping an OPO with a pulsed source such as a mode-locked optical frequency comb, as opposed to a continuous-wave beam, introduces a considerable complications. Firstly, in order to generate a nonlinear interaction, pump pulses and resonant signal pulses must overlap in the nonlinear crystal. Secondly, the OPO cavity must be dispersion-compensated for oscillating pulses, and the round-trip time for the oscillating pulses in the OPO cavity must be stabilized to match the pump repetition period. Furthermore, counter-propagating approaches require the use of complex OPO cavities containing two separate periodically-poled lithium niobate crystals[18,19]. The far simpler co-propagating geometries[18,19] have been limited by challenges in the stabilization of the carrier envelope offset (CEO), low power per comb tooth, and the observation that the spectral character of the idler pulses is highly dependent on the repetition rate[16,20].

Here we demonstrate rather unexpectedly, that a single *continuous-wave* OPO can spectrally translate a pair of *mode-locked* near-infrared femtosecond frequency combs to the mid-infrared spectral region. Thus, allowing for thousands of comb teeth to be spectrally translated with high mutual coherence and spectral bandwidths limited only by the phase matching bandwidth of the OPO. These results are enabled by: (1) chirping the pump combs such that their pulse duration is longer than the cavity round-trip time and (2) by incorporating a continuous-wave laser that is injected alongside the two combs and serves to help initiate and sustain signal oscillation. As a result, the signal beam is continuous wave, thus removing the stability challenges which limited previous work on dual-comb-pumped OPOs.

Critically, with this new approach arbitrary comb spacings can be spectrally translated and utilized for spectroscopy with total independence from the OPO cavity's free spectral range (as opposed to synchronous OPO methods which rely upon stabilization of the OPO cavity to the pump comb's repetition rate[16-19]). This method provides a facile approach for the generation of high power mid-infrared dual combs with high mutual coherence, absolute accuracy, and broad bandwidths and is expected to have wide ranging impacts in the areas of chemical kinetics, molecular spectroscopy, combustion science, and open-path remote sensing.

**Results**

The operating principle of our dual comb, continuous-wave OPO source is demonstrated in Figure 1. As described above, the use of a continuous-wave OPO removes the complexity associated with synchronous OPO while also eliminating unwanted nonlinear interactions between multiple oscillating signal beams. The two critical elements of our method are: (1) chirping of the mode-locked pump laser pulse lengths to exceed the OPO cavity length and (2) the use of a continuous-wave seed laser to initiate signal beam oscillation.

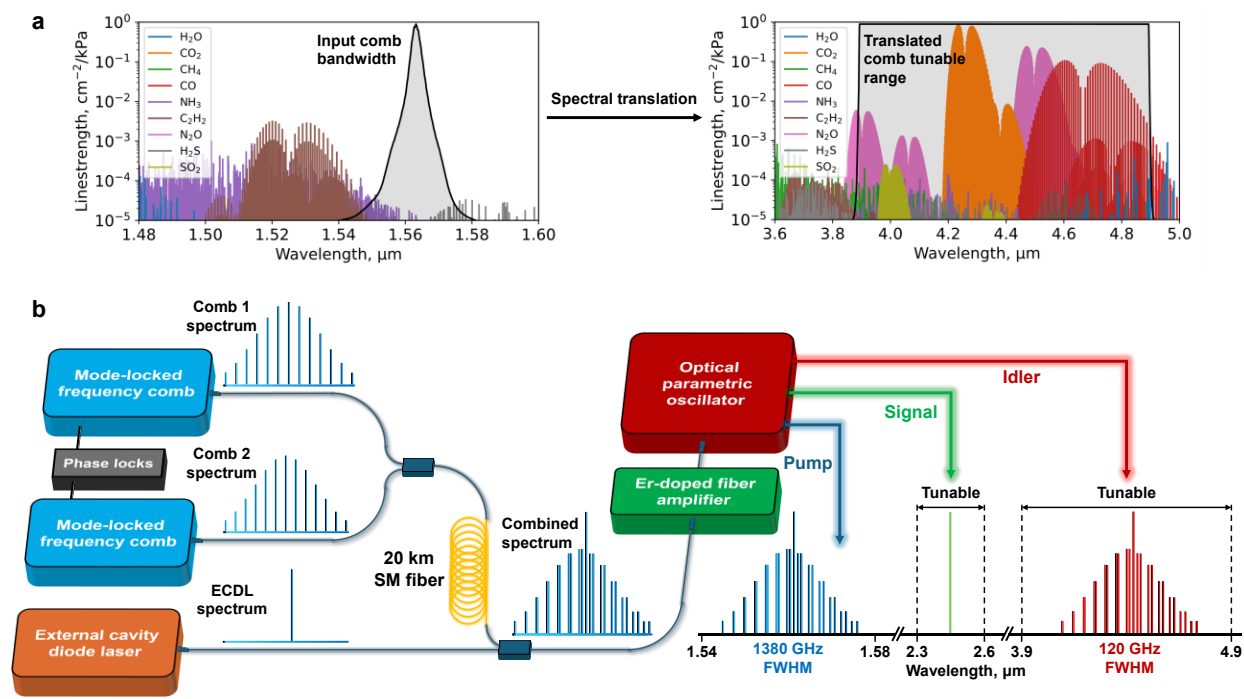

Figure 1. Concept for spectral translation of near-infrared dual combs to the mid-infrared via continuous wave optical parametric oscillation (OPO). a. Spectral translation from the near-infrared region, where molecular absorption is limited and cross-sections are small, to the mid-infrared offers significant advantages. The linestrength plots show absorption transitions of select molecules from the HITRAN 2020 database[22] with linestrengths greater than $10^{-5}$ cm$^{-2}$/kPa at 296 K in these spectral regions. b. Experimental schematic. Two near-infrared, mode-locked optical frequency combs were phase locked to one another with repetition rates which differed by 1 kHz. These two comb outputs were then combined before being passed through a 20 km single-mode (SM) fiber which served to increase the pulse durations. The output of the single-mode fiber was then combined with the output of an external-cavity diode laser (ECDL) whose wavelength was tuned to the center of the output wavelengths of the OFCs. This ECDL served to seed the signal output of the OPO. The combination of the ECDL and OFCs was then amplified to 20 W with an erbium-doped fiber amplifier before pumping the continuous wave, singly resonant OPO. The idler and depleted pump outputs of the OPO exhibited a pair of coherent OFCs (i.e., a dual comb spectrum), while the signal output was continuous wave.

We begin with a pair of femtosecond, mode-locked lasers with repetition rates near 200 MHz, center wavelengths of 1563 nm, and full-width at half maximum (FWHM) spectral bandwidths of 1230 GHz (at 3 dB, see Extended Data Figure 1). The CEOs and repetition rates of both combs were stabilized using an *f*-2*f* interferometer and phase locking to a common near-infrared diode laser. To enable dual comb spectroscopy, we configured the two optical frequency combs such that their repetition rates differed by 1 kHz. For further description of the near-infrared comb operation, see the Methods section.

By beating the two mode-locked combs directly on a photodiode we can readily observe dual comb interferograms occurring at a rate of 1 kHz (see Figure 2). Further, taking a Fourier transform of the time domain data allows us to observe the characteristic dual comb frequency domain spectra (see Figure 3). We chirped the pump mode-locked combs using 20 km of single-mode optical fiber to stretch the pump pulses to 2.5 ns duration. This ensured that the combs' pulse lengths of 75 cm are longer than the 60 cm OPO cavity length. These two combs were then combined in fiber with an external-cavity diode laser (ECDL) whose wavelength was tuned to near the center of the combs' output spectra. The combination of comb and continuous-wave ECDL light was then amplified to 20 W via an erbium-doped fiber amplifier before being injected into the OPO cavity.

The OPO in our setup was custom designed and utilized a periodically-poled lithium niobate (PPLN) crystal inside a butterfly optical cavity. The optical layout was similar to that found in Foote et al.[23], however with a designed pump wavelength of 1560 nm rather than 1060 nm. The OPO was continuous wave and singly resonant such that only the signal beam was resonant within the OPO cavity. This led to an idler tuning range from 3.9 µm to 4.9 µm, corresponding to a signal tuning range of 2.6 µm to 2.3 µm. Further details on the OPO can be found in the Methods section.

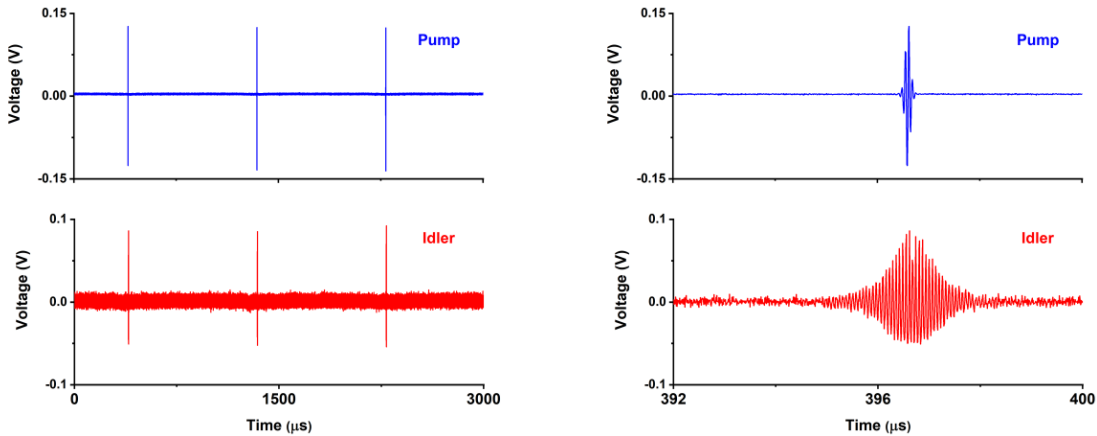

Figure 2. Time-domain interferograms for the near-infrared pump (shown in blue, upper panels) and mid-infrared idler (shown in red, lower panels) dual optical frequency combs, showing that the nature of the mode-locked pulses is preserved through the OPO. The left panel shows a series of centerbursts which occur at the repetition rate difference (1 kHz). The right panel shows a narrowed time window of a single centerburst. For these measurements the near-infrared pump dual comb was centered near 1563 nm and the mid-infrared idler dual comb was centered near 4580 nm.

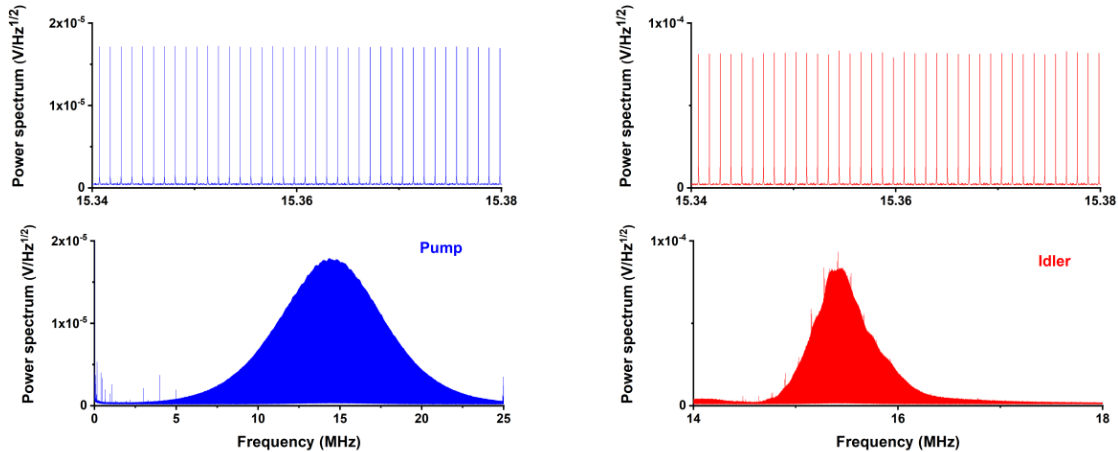

Figure 3. Frequency domain signals for the near-infrared pump (shown in blue, left panels) and mid-infrared idler (shown in red, right panels) dual optical frequency combs. The left panels show the near-infrared pump dual comb spectrum, where the upper panel is a narrowed frequency range in which the individual comb teeth can be observed. The right panels show corresponding spectra for the mid-infrared idler dual comb which contained more than a thousand comb teeth from each comb. Both dual comb spectra exhibit heterodyne beat spacings given by the difference of the near-infrared pump comb repetition rates (i.e., 1 kHz). For these measurements the near-infrared pump dual comb was centered near 1563 nm and the mid-infrared idler dual comb was centered near 4580 nm. The power spectra shown are the average of 25 individual spectra, each of which had an integration time of 95 ms.

In our method, once cavity oscillation begins, the generated signal pulses are ensured to overlap with the pump (either the incoming comb pulses or continuous-wave seed light) so that they are amplified on subsequent passes. However, highly asynchronous pump pulses generate new signal pulses with each subsequent pump pulse, and after many round-trips interference cause the signal to be continuous wave (see Extended Data Figure 2). Conversely, the pulsed character of the near-infrared pump optical frequency combs as well as their mutual coherence is preserved following spectral translation to the mid-infrared idler beam (see Figure 2). As can be seen in Figure 3, the spectrally translated, mid-infrared frequency combs give rise to characteristic dual comb heterodyne beats when placed on a photodiode and Fourier transformed. As expected, the repetition rates of the optical frequency combs are preserved through the OPO process.

In contrast to previous work on synchronous dual comb OPO[16-19], in our approach the two idler combs share a common signal beam. The presence of this strong, continuous-wave signal tone provides a convenient means of stabilizing the absolute frequency of the produced idler light. This removes one of the significant challenges which limited previous work on synchronous comb OPO; namely that difficulties in stabilizing the CEOs of the idler combs precluded the acquisition of dual comb spectra[16,24].

The observed mid-infrared dual comb spectra had a FWHM optical bandwidth near 120 GHz (see Figure 3). This width is narrower than the 1380 GHz width of the pump combs, which aligns with expectations based upon the OPO's calculated FWHM phase matching of approximately 56 GHz at this wavelength (see Extended Data Figure 3). Importantly, the use of a more tightly focused pump beam has been shown to significantly increase the OPO phase matching, with values as high as 5.7 THz demonstrated at 3.4 µm[25]. Further, we note that with oscillation at other wavelengths and in nonlinear media with broader phase matching, the bandwidth of the current approach can be greatly expanded.

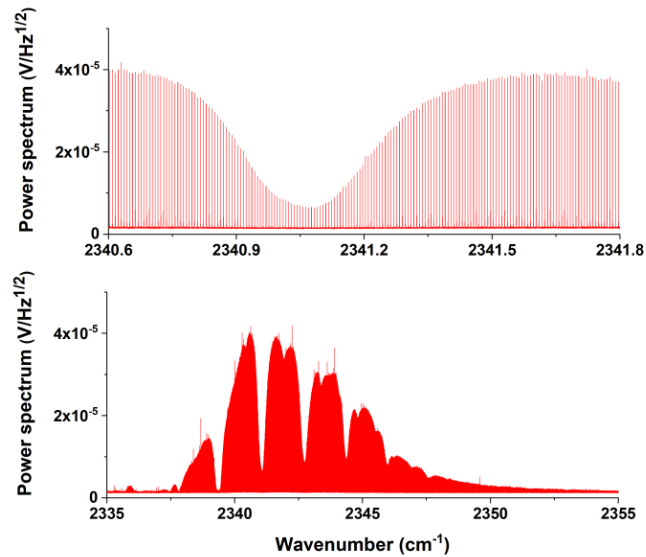

Figure 4. Mid-infrared dual comb spectrum following a 21 cm open air path (lower panel). The spectrum shown is an average of 100 individual spectra each of which was acquired over 95 ms. A series of individual carbon dioxide transitions can be observed corresponding to both strong fundamental transitions (00011←00001) and weaker hot band transitions (01111←01101). A magnified portion of the spectrum centered on an individual fundamental carbon dioxide transition can be seen in the upper panel. The spectrum encompasses nearly 3000 optical comb teeth from each comb.

Using the near-infrared mode-locked dual comb as the pump, we achieved mid-infrared idler optical powers of up to 0.4 W. This output power was similar to that observed when the OPO was pumped simply by a continuous-wave laser, indicating that the introduction of the dual optical frequency combs on the pump beam does not lead to a loss of efficiency. The OPO output power spectra indicated that sixty percent of the mid-infrared power was found in the optical frequency combs while forty percent was continuous wave (from the ECDL seed). Interestingly, in the combined, unamplified near-infrared pump light the optical frequency combs were only twenty percent of the overall power. Thus, it appears that the combination of the amplifier and optical parametric oscillation preferentially enhances the power in the combs at the expense of the continuous-wave seed light. Combining the overall mid-infrared comb powers with the power spectrum found in Figure 3, we can estimate that the present approach allows for the generation of mid-infrared combs with average powers per comb tooth at the hundreds of microwatt level. This power per comb tooth is more than two orders of magnitude higher than typical power levels[21] using other approaches such as difference frequency generation or synchronous OPO.

As can be seen in Figure 4, in the presence of open path absorption we can readily observe fourteen separate carbon dioxide transitions sampled by almost three thousand mid-infrared comb teeth in each comb. By selecting the known comb teeth frequencies in the dual comb spectrum and removing the baseline induced by the comb's spectral envelope, we can extract the spectrum shown in Figure 5. Despite the measurement time of 2.4 s, we observe good agreement between the observed spectrum and a spectrum calculated based upon the HITRAN 2020 database[22], indicating the potential of this approach for rapid, quantitative spectroscopic measurements in a multiplexed fashion.

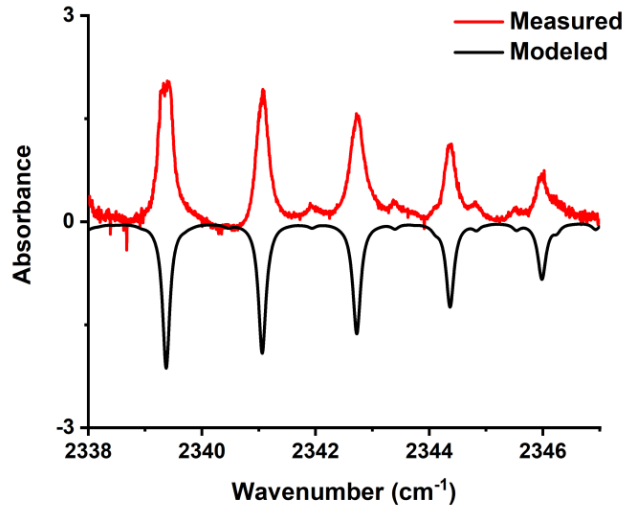

Figure 5. Background-subtracted absorbance spectrum of carbon dioxide transitions measured over a 21 cm open path. The spectrum shown was recorded with a total measurement time of 2.4 s and contains over 1300 individual comb teeth from each comb. Also shown is a simulation based upon the HITRAN 2020 database[22] produced from a fit of the absorbance spectrum where only the carbon dioxide concentration was floated (inverted for clarity).

**Conclusions and Outlook**

The present continuous-wave OPO approach offers considerable advantages over previous synchronous single OPO dual comb methods[16-19]. Firstly, the present continuous-wave approach does not require any stabilization of the pump combs to the OPO cavity and allows for complete frequency agility of the comb repetition rates. Secondly, the present approach does not rely on a complicated OPO cavity design or multiple non-linear crystals[18,19]. Thirdly, the continuous-wave signal beam provides a convenient means of monitoring and stabilizing the absolute frequency of the spectrally translated mid-infrared idler combs. Finally, as we are not relying upon high power pulses to initiate the non-linear spectral translation, the present approach is amenable to gigahertz-level repetition rates.

In the future there are several pathways by which we can expand the spectral bandwidth and optical power of the current method. Firstly, employing a pair of 1 µm mode-locked combs would allow us to utilize an OPO designed for shorter wavelength operation. This would allow for significantly broader phase matching bandwidth[26], reaching up to 5 THz FWHM which corresponds to 25 000 comb teeth. In addition, as there is far less absorption of the PPLN crystal below 4 µm we would expect idler power levels of a few Watts. Further, we would like to explore the use of gigahertz-level repetition rate combs, allowing for a unique combination of high measurement rates and wide spectral bandwidth. We note that the power of the present approach lies in its frequency agility, flexibility, and robustness with continuous-wave OPO providing a facile path for spectral translation of dual frequency comb spectroscopy to the critical mid-infrared region. We anticipate that this approach will see considerable application in combustion science, chemical kinetics and dynamics, as well as for atmospheric open-path sensing, providing a clear pathway for mid-infrared dual combs to move out of metrology laboratories and into the field.

**Methods**

Two near-infrared optical frequency combs were generated from mode-locked, Er-doped fiber lasers and used to pump the OPO. The laser oscillators were pumped with diode lasers emitting near 980 nm and generated output spectra centered near 1563 nm with FWHM bandwidths of 10.0 nm. The output of each frequency comb was split into two branches: one for phase locking and one for pumping the OPO.

Phase locking was performed by stabilizing the CEO frequency and the repetition rate of each frequency comb using digital phase lock loops. The CEO frequency was measured using a standard *f*-2*f* approach. Specifically, a small portion of the comb output was amplified with an erbium-doped fiber amplifier (EDFA), spectrally broadened in highly nonlinear fiber (HNLF) to span 1 µm to 2 µm, and passed through a fiber-coupled periodically poled potassium titanyl phosphate (PPKTP) crystal package to frequency double the low-frequency portion of the spectrum. The optical beat, which was used to stabilize the repetition rate, was generated by filtering a small portion of the spectrum with a dense wavelength division multiplexer (DWDM) and combining this with light from a narrowband continuous-wave laser near 1565 nm. The same continuous-wave laser was used to generate optical beats for both combs, creating mutual coherence between the lasers. Digital phase lock loops provided feedback to the laser pump diode current to stabilize the CEO frequency and a piezo transducer to modulate the comb's cavity length in order to stabilize the optical beat frequency of each comb. The mode-locked lasers had nominal repetition rates of 200 MHz and were stabilized with a difference in repetition rate of 1057 Hz.

The output from each comb was first combined using a polarization-maintaining (PM) fiber combiner to create a near-infrared dual comb and passed through 20 km of single-mode (SM) fiber to chirp the pulses to a duration of 2.5 ns. The dual-comb spectrum was then combined with the output of an ECDL emitting near 1563 nm using a second PM fiber combiner. This ECDL served to seed the signal oscillation of the OPO. The combined light containing the dual-comb spectrum and narrowband ECDL output was amplified to 20 W in an EDFA. The amplified light containing the dual-comb spectrum and narrowband ECDL output was then used to pump the OPO cavity[23].

The OPO was singly resonant in which only the generated signal beam was resonant in the OPO cavity. The pump was focused into the PPLN nonlinear crystal with a beam waist of approximately 60 µm. The input and output OPO cavity mirrors were curved so that the signal beam had good spatial overlap with the pump beam in the nonlinear crystal. The crystal is poled in a fan-out structure, with the poling period varying along a dimension of the crystal perpendicular to the direction of beam propagation. The PPLN crystal was mounted on a motor, which translates the crystal in this perpendicular direction in order to change the poling period exposed to the pump beam. This changes the phase-matching conditions, and thus allows wide tuning of the signal and idler. Synchronous OPOs pumped with pulsed sources benefit from high peak power in their pump pulses, providing efficient pump conversion with a short (several millimeter) nonlinear crystal. Instead, our OPO was designed to be pumped with a continuous-wave source. We used a 50 mm-long crystal to efficiently convert such a pump. The OPO cavity length was 60 cm, giving a free spectral range of 500 MHz. This is highly asynchronous with the 200 MHz pump source.

The OPO was modified from that used in previous work[23] so that it could be pumped near 1560 nm, to match the dual comb center wavelength. The resulting idler tuning curve is given in Extended Data Figure 2. The cavity mirrors were coated for high reflectivity in the new signal tuning range (2.3 µm to 2.6 µm), and for high transmission at the pump and idler range (3.9 µm to 4.9 µm). The nonlinear crystal had poling periods appropriate for the new combination of wavelengths.

Between 4.0 µm and 4.8 µm the idler power varied between 0.2 W and 1.1 W (see Extended Data Figure 4), a power which was limited by absorption in the PPLN crystal. Following spectral filtration (to remove any residual pump or signal light) the idler output of the OPO was then passed through a 21 cm open path. The light was then focused on to a HgCdTe detector having a bandwidth of 200 MHz. The resulting signal was then digitized at 125 megasamples per second by a high-speed digitizer.

It is important to note that prior experiments have shown that nonlinear effects in optical fibers can induce some spectral distortion when combs are co-propagated at high powers or over long distances in fiber[27]. Here the power at the output of the SM fiber was only 200 µW for each comb. We note that using alternative methods for chirping the comb pulses, such as fiber Bragg gratings, can further reduce any possible issues with co-propagation.

**Extended Data**

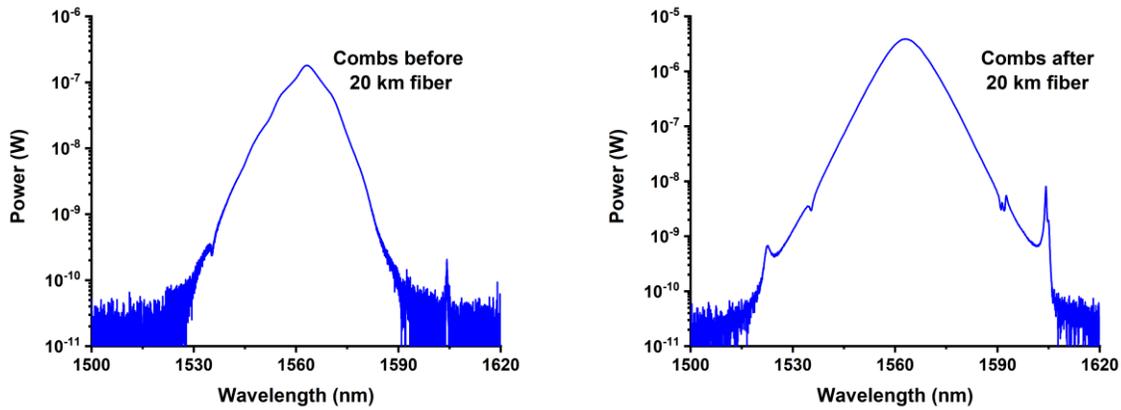

Extended Data Figure 1. Optical spectrum analyzer spectra of the near-infrared optical frequency combs both before the 20 km long single-mode fiber (left, attenuated) and after (right). The optical frequency combs exhibited FWHM spectral bandwidths of 10.0 nm (1230 GHz) and 11.3 nm (1380 GHz) before and after the 20 km fiber, respectively.

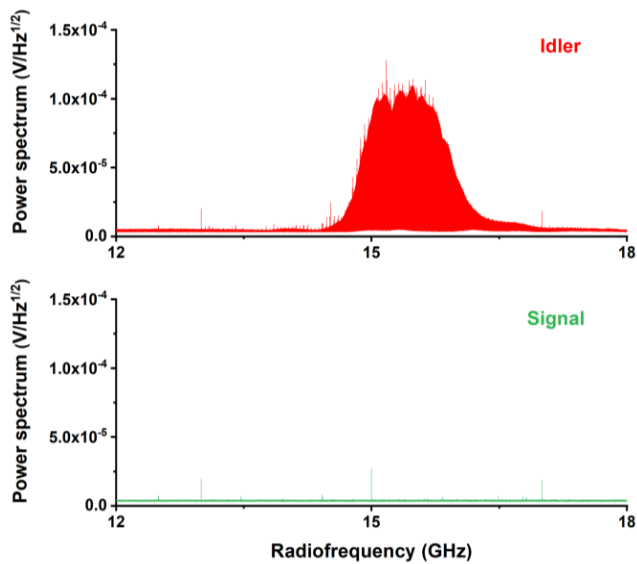

Extended Data Figure 2. Idler (upper panel in red) and signal (lower panel in green) frequency domain signals. While the idler exhibits a clear dual comb spectrum, the signal beam is continuous wave and does not show any comb behavior. The shown spectra are the average of 100 individual spectra each of which was acquired in 95 ms. These measurements were recorded when the OPO was tuned to a center idler wavelength of 4.4 µm.

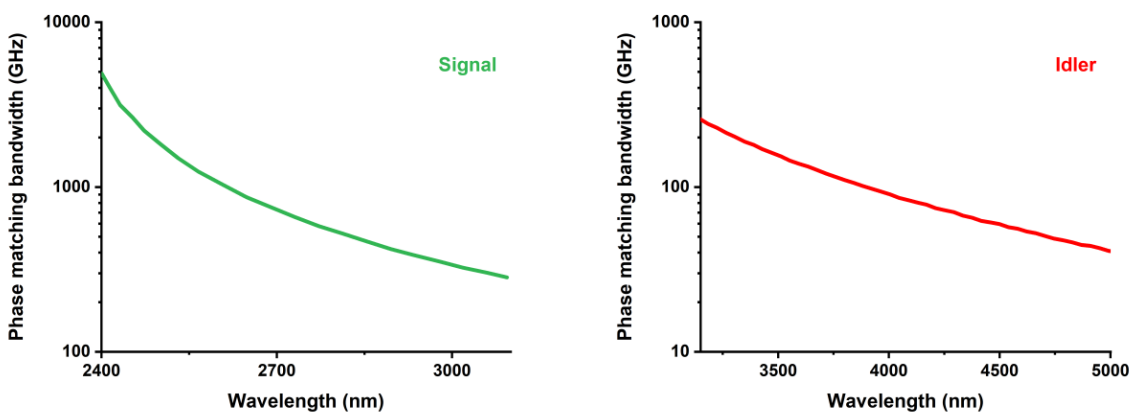

Extended Data Figure 3. Calculated FWHM phase matching bandwidth for the OPO signal (left panel in green) and idler (right panel in red).

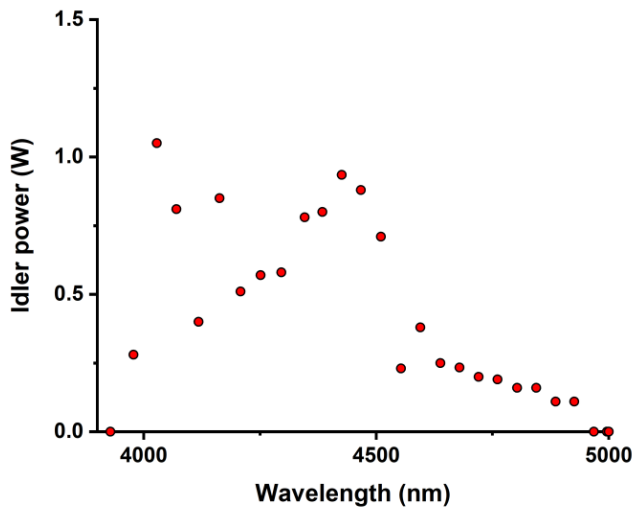

Extended Data Figure 4. Measured idler optical power as a function of wavelength.

**Acknowledgements**

We thank K. Srinivasan for helpful discussions. This material is based in part on work supported by the Air Force Office of Scientific Research under award number FA9550-20-1-0328 (G.C.M., S.C.C. and G.B.R.).